u
\documentclass[showpacs,prb,aps,amsmath]{revtex4}

\usepackage{feynmp}
\usepackage{epsf}
\usepackage[dvips]{graphicx,color,psfrag}
\usepackage{longtable}
\usepackage{amsmath}
\usepackage{bm}
\begin{document}
\title{Comments on the Chern Number Argument of the Integer Quantum Hall
Effect}

\author{Naoyuki Miyata}

\affiliation{Toyama Industrial Technology Center,\\
150, Futagami, Takaoka, Toyama, 933-0981, Japan}

\date{\today}

\begin{abstract}
The Chern number argument of the integer quantum Hall effect is
 invalid. Because a process of calculation does not change its result,
 the result of our numerical calculation means that the argument
 fails. We briefly explain why the misuse of the theory of fiber bundles happens. 
\end{abstract}
\maketitle
\section{Introduction}
Since its discovery by von Klitzing et al.\cite{klitzing}, the quantum Hall effect
(QHE) is a fascinating phenomenon for physicists, especially for those
with a close connection to fundamental physics. Among logics to explain
quantization of Hall conductance, one theory utilizes the theory of
fiber bundles\cite{kohmoto}. Although it is natural to connect Hall conductance which
is represented by integers in units of $\frac{e^2}{h}$ with the integration of
the Chern class in a cycle with integer coefficient being an integer, it
cannot be grounds for being integers.  In fact, our numerical calculation
of Hall conductance does not become an integer in units of 
$\frac{e^2}{h}$, which means the failure of the argument. A brief explanation of the weakness of the argument follows. 
\section{Model}
Our first consideration is the solution of bloch electrons in a uniform magnetic
field $B=\frac{\phi_{0}}{S}\frac{p}{q}$ where $\phi_{0}$ is the flux
quantum, S is the size of a unit cell,
and p and q are positive intergers which are mutually prime. 
The Hamiltonian is 
\begin{align}
H=\frac{1}{2m}(\hat{\bm{p}}+e\hat{\bm{A}})^{2}+V(\hat{\bm{x}})\label{hamiltonian}
\end{align}
where $m$, $e$, $\hat{\bm{p}}$, $\hat{\bm{A}}$, and $V(\hat{\bm{x}})$ are the electron
mass, the electron charge, the momentum operator, a vector potential,
and a periodic potential, respectively. We take the Landau gauge $\hat{\bm{A}}=(0,B\hat{x}_{1})$.
We present here briefly a calculation method which enables band calculation
in a magnetic field.
\\
The followings are magnetic translation operators.
\begin{align}
\tau(n_{1} \bm{a}_{1})=&e^{-\frac{i n_{1}^{2} a_{1}^{2}\sin\theta
\cos\theta}{2l^2}}e^{-\frac{i}{\hbar}(\hat{p}_{1}+\frac{\hbar}{l^2}\hat{x}_{2})n_{1}a_{1}\sin\theta}e^{-\frac{i}{\hbar}\hat{p}_{2}n_{1}a_{1}
\cos\theta}\nonumber\\
=&e^{-\frac{i\pi pn_{1}^{2}a_{1}
 \cos\theta}{a_{2}q}}e^{-\frac{i2\pi p n_{1}\hat{x}_{2}}{a_{2}q}}e^{-\frac{i}{\hbar}\hat{p}_{1}n_{1}a_{1}\sin\theta}e^{-\frac{i}{\hbar}\hat{p}_{2}n_{1}a_{1}\cos\theta},\\
\tau(n_{2}\bm{a}_{2})=&e^{-\frac{i}{\hbar}\hat{p}_{2}n_{2}a_{2}},
\end{align}
where $\bm{a}_{1}$ and $\bm{a}_{2}$ are primitive translation vectors of
a lattice, and they form an angle $\theta$. These operators commute with
a Hamiltonian $H_{0}=\frac{1}{2m}(\hat{\bm{p}}+e\hat{\bm{A}})^{2}$.
They become commutative \cite{kohmoto}when the vectors are enlarged from
$(\bm{a}_{1},\bm{a}_{2})$ to $(q\bm{a}_{1},\bm{a}_{2})$.
Then, we shall produce from the Landau level wave functions which
satisfy periodic boundary condition in terms of the magnetic
translation operators, i.e., $\tau(N_{1}
\bm{a}_{1})\psi(\bm{x})=\psi(\bm{x})$ and $\tau(N_{2}
\bm{a}_{2})\psi(\bm{x})=\psi(\bm{x})$ where $N_{1}$ and $N_{2}$ are
periodicity of the system, and $N_{1}$ is assumed to be a
multiple of $q$. On these periodic fuctions, the magnetic translation
operators form a representation of abelian magnetic translation group
(MTG) which is isomorphic to $(\bm{Z}/(N_{1}/q)\bm{Z})\times
(\bm{Z}/N_{2}\bm{Z})$.
Finally, multiplying a projection operator of an irreducible
representation of MTG onto the obtained wave functions, we obtain:
\begin{align}
& \psi_{N,\frac{2\pi}{L_{1}}n_{1},\frac{2\pi}{L_{2}}n_{2},m}(\bm{x})
 \nonumber\\=&\sqrt{\frac{q}{N_{1}}} \sum^{\frac{N_{1}}{q}}_{\tilde{n}_{1}=1}\sum^{\infty}_{\tilde{n}=-\infty}
 e^{i\frac{2\pi}{L_{1}}n_{1}\tilde{n}_{1}qa_{1}}e^{-i\pi
 \frac{pa_{1}\cos\theta}{qa_{2}}(\tilde{n}_{1}q+\tilde{n}N_{1})^{2}}e^{-i\frac{2\pi}{L_{2}}(n_{2}+N_{2}m)(\tilde{n}_{1}q+\tilde{n}N_{1})a_{1}\cos\theta}\nonumber\\
&\times \phi_{N,(-\frac{q(n_{2}+N_{2}m)}{N_{2}p}+\tilde{n}_{1}q+\tilde{n}N_{1})a_{1}\sin\theta}(x_{1})\frac{1}{\sqrt{L_{2}}}e^{i\frac{2\pi}{L_{2}}(n_{2}+N_{2}m-N_{2}(p\tilde{n}_{1}+\frac{p\tilde{n}}{q}N_{1}))x_{2}},\label{wf}
\end{align}
where $L_{1}=N_{1}a_{1}$ and
$L_{2}=N_{2}a_{2}$.
$\phi_{N,X}(x)=\frac{1}{\sqrt{2^{N}N!\sqrt{\pi}l}}(\frac{x-X}{l}-l\frac{\partial}{\partial
x})^{N}\exp(-\frac{1}{2l^{2}}(x-X)^{2})$ where $l$ is the magnetic
length, $l^2=\frac{\hbar}{eB}=\frac{\hbar}{e\frac{\phi_{0}}{S}\frac{p}{q}}=\frac{Sq}{2\pi
 p}$.

The wave function of equation (\ref{wf}) satisfies $\tau(n_{1}
q\bm{a}_{1})\psi_{N,\frac{2\pi}{L_{1}}n_{1},\frac{2\pi}{L_{2}}n_{2},m}=e^{-i\frac{2\pi}{L_{1}}n_{1}qa_{1}}\psi_{N,\frac{2\pi}{L_{1}}n_{1},\frac{2\pi}{L_{2}}n_{2},m}$, 

$\tau(n_{2}
\bm{a}_{2})\psi_{N,\frac{2\pi}{L_{1}}n_{1},\frac{2\pi}{L_{2}}n_{2},m}=e^{-i\frac{2\pi}{L_{2}}n_{1}a_{2}}\psi_{N,\frac{2\pi}{L_{1}}n_{1},\frac{2\pi}{L_{2}}n_{2},m}$.
$(k_{1},k_{2})=(\frac{2\pi}{L_{1}}n_{1},\frac{2\pi}{L_{2}}n_{2})$ are
the wave vectors in the magnetic Brillouin zone (MBZ), i.e.,
$-\frac{\pi}{qa_{1}}\leq k_{1}\leq \frac{\pi}{qa_{1}}$, $-\frac{\pi}{a_{2}}\leq k_{2}\leq \frac{\pi}{a_{2}}$.
It also satisfies
\begin{align}
 &H_{0}\psi_{N,\frac{2\pi}{L_{1}}n_{1},\frac{2\pi}{L_{2}}n_{2},m}=(N+\frac{1}{2})\hbar\omega\psi_{N,\frac{2\pi}{L_{1}}n_{1},\frac{2\pi}{L_{2}}n_{2},m}.
\end{align}
Because a periodic
potential is invariant under the magnetic translation operations,
the representation matrix in terms of the basis (\ref{wf}) is block
diagonalized. The matrix elements of a periodic potential is given in
Appendix A.

Note that inter
Landau level index $m$ varies as $0,\dots p-1$; that is, each Landau level
is p-fold degenerated and splits into p bands in general when a periodic
potential is applied. This is often misunderstood as q-fold as Douglas
Hofstadter himself did\cite{hof}.

 Although a similar method was
given by Springsguth et al.\cite{springsguth}, the reason why the method is justified is
not presented and is different from ours. Although subtlety of our method is
the periodic boundary condition in the presence of a
magnetic field, this subtlety comes from validity of using periodic
boundary condition to analyze a solid which is not periodic
actually. Moreover, people who use tight binding model with Peierls-Onsagar
substitution can not criticize our method because phase factors are the
same as them. Note that, of course, hopping integral changes in general
as magnetic field strength varies. Our method shows Hofstadter butterfly structure
as well (FIG.1). In FIG.1, as mentioned above, the p-fold degenerated lowest flat Landau
level splits into p distinct bands.
\begin{figure}[htbp]
\psfrag{vertical}{$(E-\hbar\omega/2)4\pi^{2}\epsilon_{0}/e^{2}$}
\psfrag{horizontal}{$\Phi_{0}/\Phi (=q/p)$}
\includegraphics[width=12cm,angle=270]{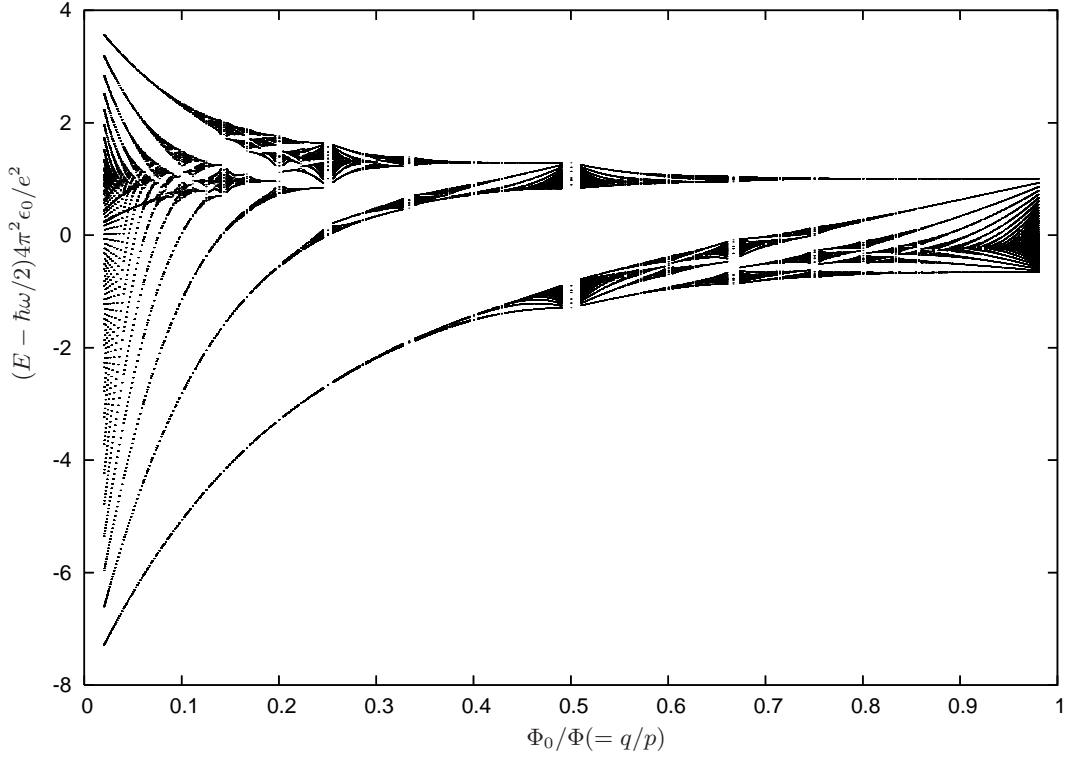}
\caption{The periodic potential is taken as
  $V(\bm{x})=\sum_{l_{1}=\pm1,l_{2}=\pm1}V_{l_{1},l_{2}}e^{2\pi
  i(\frac{l_{1}x_{1}}{a_{1}}+\frac{l_{2}x_{2}}{a_{1}})}$ where
  $V_{l_{1},l_{2}}=-\frac{e^{2}}{4\pi^{2}\epsilon_{0}}\frac{1}{l_{1}^{2}+l_{2}^{2}}$.
  The lattice is square with the
  lattice constant $1\AA$.
  Calculation was done up to $p=50$.  The horizontal axis is
  $\frac{\Phi_{0}}{\Phi}=\frac{q}{p}$. The vertical axis is
  $\frac{E-\frac{\hbar\omega}{2}}{\frac{e^{2}}{4\pi^{2}\epsilon_{0}}}$.
  It can be seen that the broadening of the lowest Landau level due to
  the periodic potential gets larger as magnetic field strength tends to infinity.}
\end{figure}  
\\

\section{Hall Conductance}
For the Hamiltonian (\ref{hamiltonian}) regardless of its potential, the Kubo
 formula for Hall conductance is

\begin{align}
 &\sigma_{12}=-\frac{\hbar e^{2}}{iVm^{2}}\sum_{n,n^{\prime},\bm{k}\in MBZ,E_{n\bm{k}}\le
 E_{F}}\frac{<n\bm{k}|\pi_{1}|n^{\prime}\bm{k}><n^{\prime}\bm{k}|\pi_{2}|n\bm{k}>-<n\bm{k}|\pi_{2}|n^{\prime}\bm{k}><n^{\prime}\bm{k}|\pi_{1}|n\bm{k}>}{(E_{n\bm{k}}-E_{n^{\prime}\bm{k}})^{2}},\label{hall}
\end{align}
where $\pi_{1}=\hat{p_{1}}, \pi_{2}=\hat{p_{2}}+eB\hat{x}_{1}$.
 When the Fermi level lies in a gap between subbands of the split
 lowest Landau level, the calculation of Hall conductance corresponds to
 the calculation,

\begin{align}
\frac{1}{2\pi i}\sum_{N,m,E_{n,\bm{k},m}\le E_{F}}\int_{MBZ}d^{2}k \int d^{2}x (\frac{\partial
 u_{N,\bm{k},m}^{\ast}}{\partial k_{2}}\frac{\partial
 u_{N,\bm{k},m}}{\partial k_{1}}-\frac{\partial
 u_{N,\bm{k},m}^{\ast}}{\partial k_{1}}\frac{\partial
 u_{N,\bm{k},m}}{\partial k_{2}})
\end{align}
where
$u_{N,\bm{k},m}^{\ast}(\bm{x})=e^{-ik_{1}x_{1}-ik_{2}x_{2}}\psi_{N,\bm{k},m}^{\ast}(\bm{x})$.
Kohmoto claims that this quantitiy is a Chern number and is therefore an
integer. To clarify the invalidity of the argument, let us calculate the
Hall conductance directly and concretely.
Equation (\ref{hall}) is transformed into 
\begin{align}
 &-\frac{ne}{B}-\frac{2me}{VB^{3}}\sum_{N,\bm{k}\in MBZ,m (E_{N,\bm{k},m}\le
 E_{F}),N^{\prime}, m^{\prime},}\frac{|<N,\bm{k},m|\frac{\partial V}{\partial x_{1}}|N^{\prime},\bm{k},m^{\prime}>|^{2}}{E_{N,\bm{k},m}-E_{N^{\prime},\bm{k},m^{\prime}}}
\end{align}
by using the commutation relation. The first term is equal to
 $-\frac{e^{2}j}{hp}$ when $j$ bands are below the Fermi energy, 
  by utilizing the number of states of MBZ of $N_{1}N_{2}/q$. Therefore, if
 the second term is negligible, the Chern number argument is invalid. In
 fact, this occurs. The result of our numerical calculation is shown in FIG.2. Plateaus
 appear as non-integers. In fact, the second term is at most of the order
 $10^{-56}$ in units of $\frac{e^{2}}{h}$, thus negligible. 

  What the author wants to say is that although Hall conductance has a form of integration of Chern class, nevertheless, it does not become an integer.
Why the integration did not become an integer is rather simple: It
forms no connection. When we take one of the largest open coverings,
i.e., whole base space minus measure zero subspace, a quantity on the
covering cannot be always regarded as a connection. In our case, this
occurred. In the argument, the basis for being integers is only that
Hall conductance is written in the same form as the Chern number, which
lacks confirmation whether the quantity made by wave functions can form
a connection. And in fact, it cannot; hence, the reason this failure
occurred.
\\

\section{Discussion}
Believers of the Chern number argument say that the Chern number within a Landau level has yet to be observed since the energy gap is too small. But can such an anomalous behavior of Hall conductance actually be observed? Experiments in a magnetic field so far show us monotonic behavior when gate voltage or magnetic field varies\cite{kawaji,willet,note}. Our calculation here is consistent with those experiments in being monotonic. Development of an experimental method that sheds light on small energy gaps will clarify which is the correct perspective.  
\\

\appendix

\section{Matrix Elements of a Periodic Potential}
Let us consider the following quantity:
\begin{align}
 & \langle
N,\frac{2\pi}{L_{1}}n_{1},\frac{2\pi}{L_{2}}n_{2},m|e^{i\frac{2\pi}{\sin\theta}(\frac{l_{1}}{a_{1}}-\frac{l_{2}\cos\theta}{a_{2}})x_{1}}e^{i2\pi\frac{l_{2}}{a_{2}}x_{2}} |N^{\prime},\frac{2\pi}{L_{1}}n_{1},\frac{2\pi}{L_{2}}n_{2},m^{\prime}\rangle.
\end{align}
Let us write $l_{2}$ as
$l_{2}=pl_{2}^{\prime}+l_{2}^{\prime\prime}$. $l_{2}^{\prime\prime}$ is
always taken to be $0\le l_{2}^{\prime\prime}\le p-1$.
We shall assume that $|l_{2}|<N_{1}^{\prime}=\frac{N_{1}}{q}$.
Then, 
(i) If $l_{2}^{\prime\prime}\ne 0$, $m^{\prime}\le
m$, and $l_{2}^{\prime}\ge 0$

\begin{align}
& \langle
N,\frac{2\pi}{L_{1}}n_{1},\frac{2\pi}{L_{2}}n_{2},m|e^{i\frac{2\pi}{\sin\theta}(\frac{l_{1}}{a_{1}}-\frac{l_{2}\cos\theta}{a_{2}})x_{1}}e^{i2\pi\frac{l_{2}}{a_{2}}x_{2}}
 |N^{\prime},\frac{2\pi}{L_{1}}n_{1},\frac{2\pi}{L_{2}}n_{2},m^{\prime}\rangle\nonumber\\
 & = \delta_{m^{\prime}+l_{2}^{\prime\prime},m}
e^{i\frac{2\pi}{L_{1}}n_{1}l_{2}^{\prime}qa_{1}}e^{-i\frac{2\pi}{L_{2}}n_{2}l_{2}^{\prime}qa_{1}\cos\theta}e^{-i\frac{\pi qa_{1}\cos\theta}{a_{2}}l_{2}^{\prime}(pl_{2}^{\prime}+2m^{\prime})}\nonumber
\\
&\times\int_{-\infty}^{\infty} dx_{1}
 \phi_{N,(-\frac{q(n_{2}+N_{2}m)}{N_{2}p})a_{1}\sin\theta}(x_{1})\phi_{N^{\prime},(-\frac{q(n_{2}+N_{2}m^{\prime})}{N_{2}p}+l_{2}^{\prime}q)a_{1}\sin\theta}(x_{1})\nonumber\\
&\times
 e^{i\frac{2\pi}{\sin\theta}(\frac{l_{1}}{a_{1}}-\frac{l_{2}\cos\theta}{a_{2}})x_{1}},
\end{align}

(ii) If $l_{2}^{\prime\prime}\ne 0$, $m^{\prime}\le
m$, and $l_{2}^{\prime}<0$,

\begin{align}
 & = \delta_{m^{\prime}+l_{2}^{\prime\prime},m}
e^{i\frac{2\pi}{L_{1}}n_{1}l_{2}^{\prime}qa_{1}}e^{-i\frac{2\pi}{L_{2}}n_{2}l_{2}^{\prime}qa_{1}\cos\theta}e^{i\frac{\pi qa_{1}\cos\theta}{a_{2}}l_{2}^{\prime}(pl_{2}^{\prime}-2m)}\nonumber\\
&\times\int_{-\infty}^{\infty} dx_{1}
 \phi_{N,(-\frac{q(n_{2}+N_{2}m)}{N_{2}p}-l_{2}^{\prime}q)a_{1}\sin\theta}(x_{1})\phi_{N^{\prime},(-\frac{q(n_{2}+N_{2}m^{\prime})}{N_{2}p})a_{1}\sin\theta}(x_{1})\nonumber\\
&\times
 e^{i\frac{2\pi}{\sin\theta}(\frac{l_{1}}{a_{1}}-\frac{l_{2}\cos\theta}{a_{2}})x_{1}},
\end{align}

(iii)  If $l_{2}^{\prime\prime}\ne 0$, $m^{\prime}\ge
m$, and $l_{2}^{\prime}+1\ge 0$,

\begin{align}
 & = \delta_{m^{\prime}+l_{2}^{\prime\prime},m+p}
e^{i\frac{2\pi}{L_{1}}n_{1}(l_{2}^{\prime}+1)qa_{1}}e^{-i\frac{2\pi}{L_{2}}n_{2}(l_{2}^{\prime}+1)qa_{1}\cos\theta}e^{-i\frac{\pi qa_{1}\cos\theta}{a_{2}}(l_{2}^{\prime}+1)(p(l_{2}^{\prime}+1)+2m^{\prime})}\nonumber
\\
&\times\int_{-\infty}^{\infty} dx_{1}
 \phi_{N,(-\frac{q(n_{2}+N_{2}m)}{N_{2}p})a_{1}\sin\theta}(x_{1})\phi_{N^{\prime},(-\frac{q(n_{2}+N_{2}m^{\prime})}{N_{2}p}+(l_{2}^{\prime}+1)q)a_{1}\sin\theta}(x_{1})\nonumber\\
&\times
 e^{i\frac{2\pi}{\sin\theta}(\frac{l_{1}}{a_{1}}-\frac{l_{2}\cos\theta}{a_{2}})x_{1}},
\end{align}

(iv) If $l_{2}^{\prime\prime}\ne 0$, $m^{\prime}\ge
m$, and $l_{2}^{\prime}+1< 0$,

\begin{align}
 & = \delta_{m^{\prime}+l_{2}^{\prime\prime},m+p}
e^{i\frac{2\pi}{L_{1}}n_{1}(l_{2}^{\prime}+1)qa_{1}}e^{-i\frac{2\pi}{L_{2}}n_{2}(l_{2}^{\prime}+1)qa_{1}\cos\theta}e^{i\frac{\pi qa_{1}\cos\theta}{a_{2}}(l_{2}^{\prime}+1)(p(l_{2}^{\prime}+1)-2m)}
\nonumber\\
&\times\int_{-\infty}^{\infty} dx_{1}
 \phi_{N,(-\frac{q(n_{2}+N_{2}m)}{N_{2}p}-(l_{2}^{\prime}+1)q)a_{1}\sin\theta}(x_{1})\phi_{N^{\prime},(-\frac{q(n_{2}+N_{2}m^{\prime})}{N_{2}p})a_{1}\sin\theta}(x_{1})\nonumber\\
&\times
 e^{i\frac{2\pi}{\sin\theta}(\frac{l_{1}}{a_{1}}-\frac{l_{2}\cos\theta}{a_{2}})x_{1}},
\end{align}

(v) If $l_{2}^{\prime\prime}=0$ and $l_{2}^{\prime}\ge 0$,

\begin{align}
 & = \delta_{m^{\prime}+l_{2}^{\prime\prime},m}
e^{i\frac{2\pi}{L_{1}}n_{1}l_{2}^{\prime}qa_{1}}e^{-i\frac{2\pi}{L_{2}}n_{2}l_{2}^{\prime}qa_{1}\cos\theta}e^{-i\frac{\pi qa_{1}\cos\theta}{a_{2}}l_{2}^{\prime}(pl_{2}^{\prime}+2m^{\prime})}
\nonumber\\
&\times\int_{-\infty}^{\infty} dx_{1}
 \phi_{N,(-\frac{q(n_{2}+N_{2}m)}{N_{2}p})a_{1}\sin\theta}(x_{1})\phi_{\tilde{N},(-\frac{q(n_{2}+N_{2}m^{\prime})}{N_{2}p}+l_{2}^{\prime}q)a_{1}\sin\theta}(x_{1})\nonumber\\
&\times
 e^{i\frac{2\pi}{\sin\theta}(\frac{l_{1}}{a_{1}}-\frac{l_{2}\cos\theta}{a_{2}})x_{1}},
\end{align}

(vi) If $l_{2}^{\prime\prime}=0$ and $l_{2}^{\prime}< 0$,

\begin{align}
 & = \delta_{m^{\prime}+l_{2}^{\prime\prime},m}
e^{i\frac{2\pi}{L_{1}}n_{1}l_{2}^{\prime}qa_{1}}e^{-i\frac{2\pi}{L_{2}}n_{2}l_{2}^{\prime}qa_{1}\cos\theta}e^{i\frac{\pi qa_{1}\cos\theta}{a_{2}}l_{2}^{\prime}(pl_{2}^{\prime}-2m)}\nonumber\\
&\times\int_{-\infty}^{\infty} dx_{1}
 \phi_{N,(-\frac{q(n_{2}+N_{2}m)}{N_{2}p}-l_{2}^{\prime}q)a_{1}\sin\theta}(x_{1})\phi_{N^{\prime},(-\frac{q(n_{2}+N_{2}m^{\prime})}{N_{2}p})a_{1}\sin\theta}(x_{1})\nonumber\\
&\times
 e^{i\frac{2\pi}{\sin\theta}(\frac{l_{1}}{a_{1}}-\frac{l_{2}\cos\theta}{a_{2}})x_{1}}.
\end{align}
\begin{figure}[htbp]
 \begin{center}
\psfrag{y}{$\sigma_{12}h/e^{2}$}
\psfrag{x}{$(E_{F}-\hbar\omega/2)4\pi^{2}\epsilon_{0}/e^{2}$}
\includegraphics[width=12cm,angle=270,origin=c]{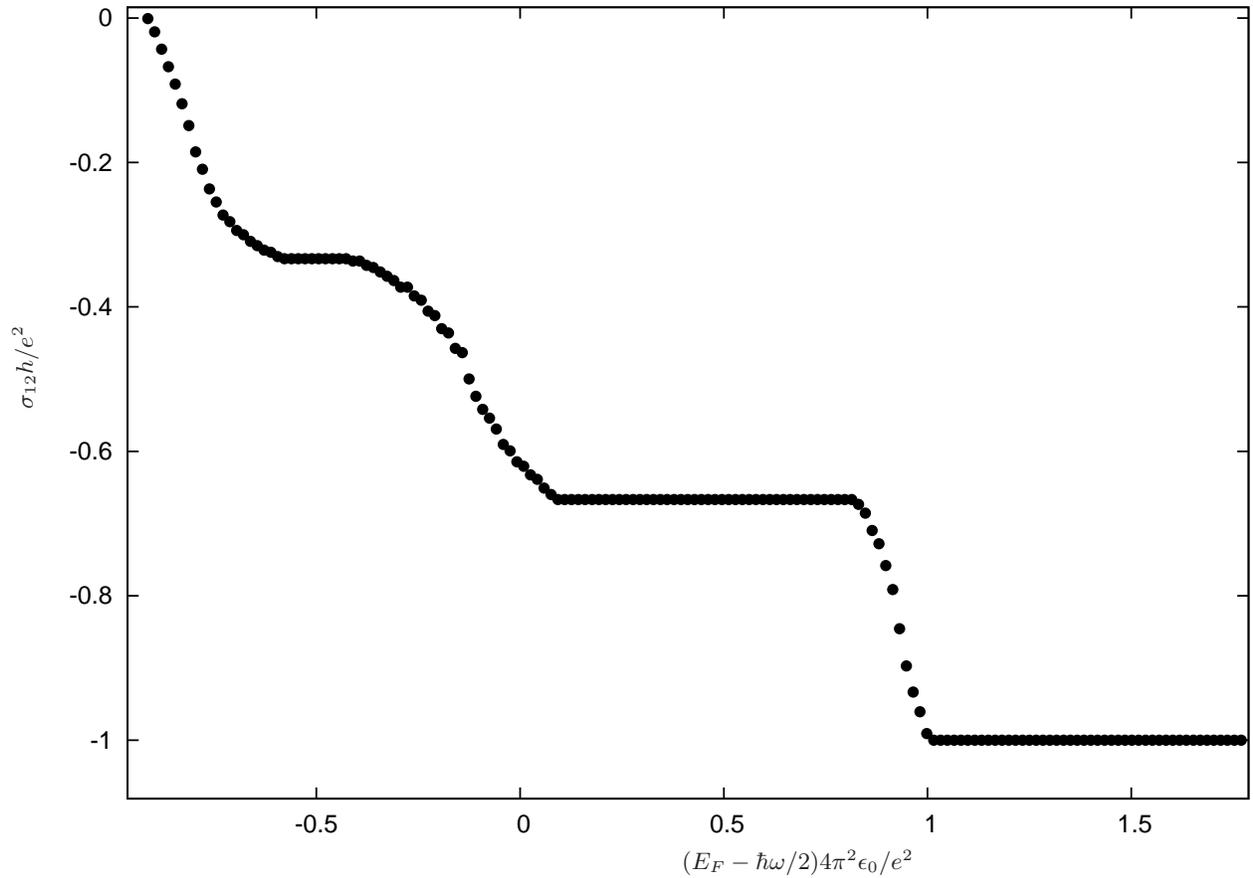}
\caption{The horizontal axis is
  $\frac{(E_{F}-\hbar\omega/2)}{\frac{e^{2}}{4\pi^{2}\epsilon_{0}}}$, where $E_{F}$ is the
  Fermi energy. The vertical axis is the Hall conductance in units of
  $\frac{e^{2}}{h}$. The condition is the same as FIG.1. The calculation
  is done at $\frac{q}{p}=\frac{2}{3}$. At $\frac{q}{p}=\frac{2}{3}$, the lowest Landau
  level splits into three bands as can be seen in FIG. 1. When the Fermi
  energy is in the gaps, the Hall conductance exhibits plateaus of
  approximately $-\frac{1}{3},-\frac{2}{3},-1$.} 
 \end{center}
\end{figure}
\\


\begin{thebibliography}{}
\bibitem{note} We do not take into account the Rashba term.
\end{thebibliography}
\end{document}